\documentclass[smallextended, referee, epjc3]{svjour3}

\RequirePackage[T1]{fontenc}

\smartqed  

\RequirePackage{graphicx}
\RequirePackage{mathptmx}      
\RequirePackage{flushend}
\RequirePackage[numbers,sort&compress]{natbib}
\RequirePackage[colorlinks,citecolor=blue,urlcolor=blue,linkcolor=blue]{hyperref}

\newcommand\data{x^{\rm d}}
\def\omms{\Omega_m}

\journalname{Eur. Phys. J. C}

\begin{document}
\title{Observational tests of the Glavan, Prokopec and Starobinsky model of dark energy}

\author{Marek Demianski\thanksref{e1,addr1,addr2}
        \and
        Ester Piedipalumbo\thanksref{e2,addr3,addr4} 
}

\thankstext{e1}{e-mail: mde@fuw.edu.pl}
\thankstext{e2}{e-mail: ester@na.infn.it}

\institute{Institute for Theoretical Physics, University of Warsaw, Pasteura 5, 02-093 Warsaw, Poland,\label{addr1}
          \and
          Department of Astronomy, Williams College, Williamstown, MA 01267, USA,\label{addr2}
         \and  Dipartimento di Fisica, Universit\`{a} degli Studi di Napoli Federico II, Compl. Univ. Monte S. Angelo, 80126 Naples, Italy,\label{addr3}
          \and
          I.N.F.N., Sez. di Napoli, Compl. Univ. Monte S. Angelo, Edificio 6, via Cinthia, 80126 - Napoli, Italy. \label{addr4}
          }

\date{Received: date / Accepted: date}

\maketitle

\begin{abstract}
In the last dozens of  years different data sets revealed the accelerated expansion of the Universe
which is driven by the so called dark energy, that now dominates the total amount of matter-energy in the Universe.
In a recent paper Glavan,  Prokopec  and  Starobinsky propose an interesting model of dark energy,
which traces the Universe evolution from the very early quantum era to the present time.
Here we perform a high-redshift analysis
to check if this new model is compatible with present day observational data and compare predictions of this model
with that of the standard $\Lambda$CDM cosmological model.
In our analysis we use only the most reliable observational data, namely
the distances to selected SNIa,  GRB Hubble diagram,
and $28$ direct measurements of the Hubble constant. Moreover we consider also  non  {\it geometric} data related to
the growth rate of density perturbations. We explore the probability distributions of the cosmological parameters
for both models. To build up their own regions of confidence,  we maximize the appropriate likelihood
functions using the Markov chain Monte Carlo (MCMC) method. Our statistical analysis indicates that  these  very
different models of dark energy are compatible with present day observational data and the GPS model seems slightly
favored  with respect to the  $\Lambda$CDM model. However to further restrict
different models of dark energy it will be necessary to increase the precision of the Hubble diagram at high redshifts
and  to perform more detailed analysis of the influence of dark energy on the process of formation of large scale structure.
\end{abstract}

\date{}


\section{Introduction}
\label{sec:Introduction}
The discovery in late 1990s that expansion of the Universe is accelerating \cite{Riess98}, \cite{Perlmutter99}
strengthened the conviction that
the Universe is spatially flat by making the total mass-energy density parameter $\Omega_{tot} = 1$
and it prompted cosmologists and physicists to ask questions about the nature of
the medium that is causing this acceleration. Now it is called dark energy and is usually assumed to uniformly fill the Universe. One
possible candidate for dark energy is the cosmological constant introduced by Einstein in 1917 when he
proposed the first static cosmological model based on general theory of relativity. In 1968 Zeldovich
\cite{Zeldovich68} noticed that properties of the quantum vacuum energy density mimic properties of the cosmological
constant. But estimates of the quantum vacuum energy density by many orders
of magnitude exceed the observational limits on energy density of dark energy. It soon turned out
that accelerated expansion of the Universe can be driven by potential energy of some evolving
self interacting scalar field. There are also other possibilities listed and discussed in
the recent review by Joyce, Lombriser and Schmidt \cite{Joyce2016}. In hydrodynamical approximation dark energy
is represented as a medium characterized by energy density $ \varrho_{DE}$ and pressure $ p_{DE}$
that are related by a simple equation of state
\begin{equation} p_{DE} = w \varrho_{DE}\,,
\end{equation}
where the proportionality coefficient $w$, in general, could
depend on time. To generate accelerated
 expansion $w < - 1/3$, $ w= -1$ for the cosmological constant.
How dark energy is influencing the expansion rate of the Universe is dictated by the mass-energy
conservation laws of different constituents and the Friedman equation. Let us assume that the Universe
is spatially flat and is described by the FLRW metric
\begin{equation}
ds^2 = dt^2 - a^{2}(t)\big[ dr^2 + r^2( {d\theta}^2 + \sin^{2}{\theta}{d\varphi}^2) \big]\,.\end{equation}
The mass-energy conservation laws for the basic non interacting
constituents are
\begin{equation} {\dot \varrho}_{r} + 4H \varrho_{r} = 0\,,\end{equation}
\begin{equation} {\dot \varrho}_{m} + 3H \varrho_{m} = 0\,,\end{equation}
\begin{equation} {\dot \varrho}_{DE} + 3H \varrho_{DE}(1 + w) = 0\,,\end{equation}
and the Friedman equation is
\begin{equation}
H^2(a)= \big( {{\dot a}\over a} \big)^2= {{8\pi G}\over 3} \big( \varrho_{r}(a) + \varrho_{m}(a) + \varrho_{DE}(a)
\big)\,,
\end{equation}
where $\varrho_{r}$ is the density of radiation (photons and other relativistic particles), $\varrho_{m}$ is the density of matter
(baryons and dark matter), $\varrho_{DE}$ is the density of dark energy and $a$ is the cosmological scale factor. Integrating the conservation laws we get:
\begin{equation}
\varrho_{r}(a) = \varrho_{r}(0)\big( {a_0 \over a}\big)^4\,,
\end{equation}
\begin{equation}
\varrho_{m}(a) = \varrho_{m}(0)\big( {a_0 \over a}\big)^3\,,
\end{equation}
when $ w = -1$ (cosmological constant) $\varrho_{DE} = {\rm const}$, when $ w \not= -1$ but it is constant
\begin{equation}
\varrho_{DE}(a) = \varrho_{DE}(0)\big( {a_0 \over a}\big)^{3(1 + w)}\,,
\end{equation}
and when $w$ is time dependent
\begin{equation} \varrho_{DE}(a) = \varrho_{DE}(0)\bigl( {a_0 \over a}\bigr)^3 \exp{\bigl( -3\int_{a_{0}}^{a}{ { w(x) \over x}dx}}\bigr)\,.
\end{equation}
The Friedman equation can be conveniently rewritten in the form
\begin{eqnarray}
H^{2}(z)& =& H^{2}(0)\big( \Omega_{r}(1 + z)^4 + \Omega_{m}(1 +z)^3 + \Omega_{DE}(1 + z)^3\nonumber\\&& \exp{ \big( 3\int_{0}^{z}{ {w(x) \over { 1 + x}}dx}\big)}\big)\,,
\end{eqnarray}
where $z$ is the redshift parameter normalized so that $a =
\displaystyle{{1 \over {1 +z}}}$, $H(0)$ is the present value of
the Hubble constant and $ \Omega_{i} =
\displaystyle{{\varrho_{i} \over \varrho_{crit}}}$ are the
density parameters of different constituents of the Universe and
$\varrho_{crit} =\displaystyle{ {{3 H^{2}(0)}\over {8 \pi G}}}$.
Different models of dark energy predict different dependence of
$w$ on the redshift $z$.
\section{ The Glavan, Prokopec, Starobinsky (GPS) model of dark energy}
In the recent paper "Stochastic dark energy from inflationary
quantum fluctuations" based on their earlier ideas and
calculations  Glavan,  Prokopec  and
Starobinsky \cite{Glavan017} propose an interesting model of dark energy that, in
what follows, we will call the GPS model. They consider a very
light non minimally coupled spectator scalar field and trace its
evolution from the very early quantum era to the present time.
They consider  the spatially flat
Friedman-Lemaitre-Robertson-Walker background spacetime with the metric
$$ds^2 = - dt^2 + a^{2}(t)\big( dx^2 + dy^2 + dz^2 \big)\,,$$
where $a(t)$ is the scale factor that satisfies the Friedman equations
\begin{equation} H^{2}(t)= \big( {{\dot a}\over a}\big)^2= {1\over {3M_{Pl}^2}}\varrho_{c}\,,
\end{equation}
\begin{equation}{\dot H} = - {1\over {2M_{Pl}^2}}(\varrho_{c} + p_{c})\,,\end{equation}

where $M_{Pl}= (8\pi G)^{-1/2}$ is the reduced Planck mass, $G$
is the Newton's gravitational constant, and $\varrho_{c}$ and
$p_{c}$ are the energy density and pressure of the dominant
cosmological constituent treated as a classical fluid.

Evolution of the non minimally coupled scalar field is determined
by the action
\begin{eqnarray} S[\Phi] &=&\int{d^4x {\cal L}_{\Phi}} =\\&& \int{ d^4x\sqrt{-g}\big( -{1\over
2}g^{\mu\nu}\partial_{\mu}\Phi\partial_{\nu}\Phi - {1\over
2}m^2\Phi^{2} - {1\over 2}\xi R \Phi^2\big)}\nonumber\,,\end{eqnarray} where $m$ is
mass of the scalar field, $R$ is the Ricci scalar and $\xi$ is
the non minimal coupling constat parameter.

They quantize the scalar field using the standard procedure of
canonical formalism and they study evolution of this quantum
scalar field and its influence on the background geometry. They
show that the vacuum expectation values of the energy-momentum
tensor operator of the quantum scalar field is diagonal and
represents an ideal fluid with energy density $\varrho_{Q}(t)$
and pressure $p_{Q}(t)$. To study the back reaction of quantum
vacuum fluctuations they decompose the field operators into long
and short wavelength modes and concentrate on evolution of the
long wavelength modes only. It turns out that the evolution
equations of the long wavelength modes contain source terms that
originated from the coupling between the short and long
wavelength modes. The source terms can be considered as
stochastic forces acting on the long wavelength modes. Next they
derive equations of motion for appropriately normalized equal
time two point correlation functions $\Delta_{\phi\phi}(t)$,
$\Delta_{\phi\pi}(t)$, and $\Delta_{\pi\pi}(t)$.

The expectation values of the energy density and pressure can be
expressed by the equal time correlation functions as
\begin{equation} \varrho_{Q} \approx {H^2\over 2}\big( \Delta_{\pi\pi} +
6\xi\Delta_{\phi\pi}+ [ 6\xi + \big( {m\over
H}\big)^2]\Delta_{\phi\phi}\big)\,,\end{equation}
\begin{eqnarray}
p_{Q}& \approx &{H^2\over 2}\big( ( 1 - 4\xi)\Delta_{\pi\pi} +
2\xi\Delta_{\phi\pi} \\&+& [ - 2\xi(3 - 2\varepsilon) + 24\xi^2(2 -
\varepsilon)- ({m\over H})^2(1 -
4\xi)\Delta_{\phi\phi}]\big)\nonumber\,,\end{eqnarray}

where $\varepsilon = -\displaystyle{{{\dot H}\over H^2}}$ is a
parameter that characterizes the equation of state of the medium
that dominates the expansion rate of the Universe, so
$\varepsilon \sim 0$ during the inflation epoch, $\varepsilon =
2$ during radiation dominated epoch and $\varepsilon = {3\over
2}$ during matter dominated epoch.

Later they study the quantum induced corrections $\varrho_{Q}$
and $ p_{Q} $ at the early period of inflation, radiation
dominated period and matter dominated period. Finally they
consider the late stage of evolution of the Universe that begins at
an arbitrarily set initial moment $z_{in}=10$, at that moment
$t=t_{*}$, $H(t_{*}) = H_{*} >> H_{DE}$, where $H_{DE}$ is the
value of the Hubble expansion rate at the beginning of the epoch
of dark energy domination.

To study evolution of different cosmological parameters it is more convenient to use instead of time $t$ or redshift
$z$ the number of e-foldings $N=  \ln( {a_{0}\over a})$. It turns out that the quantum backreaction becomes essential
when the model parameters satisfy the following conditions:
$$({m\over H})^2 << 1 \,, \quad N_{I} < {1\over
{8|\xi|}}\big[4\pi({M_{Pl}\over H_{I}})^2\big]\,, \quad 0 > \xi > -{1\over 6}({m\over H_{DE}})^2\,,$$
where $H_{I}$ is Hubble's parameter during the
period of inflation and $N_{I}= \displaystyle{{1\over
{8|\xi|}}\ln \big[ 24 \pi |\xi|({M_{Pl}\over
H_{I}})^2({H_{DE}\over m})^2\big]}\,.$

Unfortunately the evolution equations at the late stage can be
solved only numerically. Numerical results suggest the following
parametrization of the dark energy equation of state
\begin{equation} p_{Q}= w_{Q}(N) \varrho_{Q}\,,\end{equation}
where
\begin{equation} w_{Q}(N) = - {\omega_{0}\over 2}\big[ 1 + \tanh({{N -
n_{0}}\over {\delta n}})\big] = - {\omega_{0}\over {1 + \beta (1
+ z)^\alpha }}\,,
\end{equation}
where $\omega_{0} \leq 1$, $n_{0} = \displaystyle{{{\ln(\beta)}\over \alpha}}$ and $\delta n = \displaystyle{2\over
\alpha}$.

\section{ Observational tests of the GPS dark energy}
At the late stage of evolution, at $ z \leq 10$, the Universe is filled in with dark matter, baryonic matter and dark
energy. Dark matter is usually assumed to be cold and collisionless so both types of matter could be treated as
pressureless dust with mass-energy density $\varrho_{m}$. Dark matter and baryonic mater does not interact with dark
energy, so both matter components and dark energy could be treated as non interacting perfect fluids. The continuity
equation for matter in the FLRW model has the simple form
\begin{equation}{\dot \varrho}_{m} + 3 H \varrho_{m}=0\,,\end{equation}
corresponding equation for the GPS dark energy is
\begin{equation}{\dot \varrho}_{DE} + 3H( 1 - {{\omega_{0} a^{\alpha}}\over
{\beta + a^{\alpha}}})\varrho_{DE}=0\,,
\end{equation}
where $H=\displaystyle{{ \dot a} \over a}$ and $\omega_{0}$\,, $\alpha$\,, $\beta$ are constants. Integrating both equations and using the
standard relation $a(z) = \displaystyle{1\over {1 + z}}$, we get
\begin{equation}\varrho_{m}(z) = \varrho_{M}(0)(1 + z)^3\,,\end{equation}
\begin{equation}\varrho_{DE} = \varrho_{DE}(0)(1 + z)^{3(1 - \omega_{0})}\big(
{{ 1 + \beta(1 + z)^{\alpha}}\over {1 +
\beta}}\big)^{3{\omega_{0}\over \alpha}}\,,\end{equation} where
$\varrho_{m}(0)$ is the present density of matter and
$\varrho_{DE}(0)$ is the present density of dark energy. The
Hubble expansion rate is
\begin{eqnarray}
H^{2}(z) &=& {{8\pi G}\over 3}\left(\varrho_{m}(0)(1 + z)^3 +
\varrho_{DE}(0)(1 + z)^{3(1 - \omega_{0})}\right.\nonumber\\&\,&\left. \left( {{ 1 + \beta(1 +
z)^{\alpha}}\over {1 + \beta}}\right)^{3{\omega_{0}\over
\alpha}}\right)\,,\end{eqnarray}
or using the density parameters $\Omega_{m}(0)$ and
$\Omega_{DE}(0)$, we get
\begin{eqnarray}\label{Hz}
H^{2}(z) &=& H^{2}(0)\left(\Omega_{m}(0)(1 + z)^3 +
\Omega_{DE}(0)(1 + z)^{3(1 - \omega_{0})}\right.\nonumber\\&\,&\left. \left( {{ 1 + \beta(1 +
z)^{\alpha}}\over {1 + \beta}}\right)^{3{\omega_{0}\over
\alpha}}\right)\,,\end{eqnarray}
We use the Hubble expansion rate to define the luminosity distance $d_{L}$, the angular diameter distance $d_{A}$ and
the volume distance $d_{V}$ as
\begin{eqnarray}&&d_L(z,{\mathrm \theta}) ={{c}\over{H_0}} (1+z)\int^{z}_{0}{1\over H(\zeta, \theta)}d\zeta\,\\
&&={{c}\over {H_0}} (1+z)\,\nonumber\\&&\int_0^z \frac{dy}{ \sqrt{\Omega_m (y+1)^3 + \Omega_{DE}(1 + y)^{3(1 - \omega_{0})}\big( {{ 1
+ \beta(1 + y)^{\alpha}}\over {1 + \beta}}\big)^{3{\omega_{0}\over \alpha}}}}\nonumber
\end{eqnarray}
\begin{eqnarray}&&d_A(z,{\mathrm \theta}) ={{c}\over{H_0}}{{1}\over{1+z}}\,\\&&\int_0^z \frac{dy}{  \sqrt{{\Omega_m (y+1)^3 + \Omega_{DE}(1 + y)^{3(1 - \omega_{0})}\big( {{ 1 +
\beta(1 + y)^{\alpha}}\over {1 + \beta}}\big)^{3{\omega_{0}\over \alpha}}}}}\nonumber\,,\end{eqnarray}
\begin{eqnarray}d_V(z,{\mathrm \theta}) &=&\left[\left(1+z\right) d_A(z,{\mathrm \theta})^2\frac{c z}{H(z,{\mathrm
\theta})}\right]^{{1}\over{3}}\,,\end{eqnarray}
where $\theta$ denote parameters of the GPS dark energy. Using
the luminosity distance, we can evaluate the distance modulus
from its standard definition
\begin{equation}\label{modulus}
 \mu(z) = 25 + 5 \log{d_L(z,{\mathrm \theta})}\,.
 \end{equation}
\section{Observational data}
In our analysis we  mainly use the same data sets that we used in our previous cosmographic analysis \cite{MGRB2}:
measurements of SNIa distances and GRB Hubble diagram,  and  $28$ direct measurements of $H(z)$
compiled in \cite{farooqb}. However, in order to address the problem of degeneracy in the dark energy
sector, that manifests in the fact that different models of dark energy are compatible with {\it geometric tests}, that
are sensitive only to the background expansion of the universe, we consider also additional observational data, which
are non {\it geometric}. Actually we use data related to the growth rate of matter density perturbations.

\subsection{Supernovae and GRB Hubble diagram}
\subsubsection{Supernovae Ia}
Observations of  SNIa gave the first strong indication that now
expansion of the Universe is accelerating. First results of the
SNIa teams were published by \cite{Riess98} and
\cite{Perlmutter99}. Here we consider the recently updated
Supernovae Cosmology Project Union 2.1 compilation
\cite{Union2.1}, which is an extension of the original Union
compilation and contains $580$ SNIa, spanning the redshift range
($0.015 \le z \le 1.4$). We compare the theoretically predicted
distance modulus $\mu(z)$ with the observed one using a Bayesian
approach, based on the definition of the distance modulus in
different cosmological models:
\begin{equation}\label{modulus2}
\mu(z_{j}) = 5 \log_{10} { d_{L}(z_{j}, \theta_{i})} )+\mu_0\,,
\end{equation}
where $\mu_{0}$ encodes the Hubble constant and the absolute magnitude $M$ and $\theta_{i}$ are model parameters. Actually, it is well known
 that using only SNIa, one cannot constrain $H_0$, without including measurements of the local value from the SHOES project \cite{shoes, shoes2}, since
 this is degenerate with $M$.  However we can indirectly estimate the Hubble constant,  joining SNIa data with other probes.  For this purpose  in  Sec. (\ref{stat}) we  introduce a gaussian prior for $H_0$ using its value determined by the SH0ES project. Given
the heterogeneous origin of the Union data set, we use an alternative version of the $\chi^2$ test
\begin{equation}
\label{eq: sn_chi_mod}
\tilde{\chi}^{2}_{\mathrm{SN}}(\theta_{i}) = c_{1} -
\frac{c^{2}_{2}}{c_{3}}\,,
\end{equation}
where
\begin{equation}
c_{1} = \sum^{{\cal{N}}_{SNIa}}_{j = 1} \frac{(\mu(z_{j};
\mu_{0}=0, \theta_{i}) -
\mu_{obs}(z_{j}))^{2}}{\sigma^{2}_{\mathrm{\mu},j}}\, ,
\end{equation}
\begin{equation}
c_{2} = \sum^{{\cal{N}}_{SNIa}}_{j = 1} \frac{(\mu(z_{j};
\mu_{0}=0, \theta_{i}) -
\mu_{obs}(z_{j}))}{\sigma^{2}_{\mathrm{\mu},j}}\, ,
\end{equation}
\begin{equation}
c_{3} = \sum^{{\cal{N}}_{SNIa}}_{j = 1}
\frac{1}{\sigma^{2}_{\mathrm{\mu},j}}\,.
\end{equation}
It is worth noting that
\begin{equation}
\chi^{2}_{\mathrm{SN}}(\mu_{0}, \theta_{i}) = c_{1} - 2 c_{2}
\mu_{0} + c_{3} \mu^{2}_{0} \,,
\end{equation}

which clearly becomes minimal for $\mu_{0} = c_{2}/c_{3}$, so
 that $\tilde{\chi}^{2}_{\mathrm{SN}} \equiv
\chi^{2}_{\mathrm{SN}}(\mu_{0} = c_{2}/c_{3}, \theta_{i})$.
\subsubsection{Gamma-Ray Burst Hubble diagram}
\begin{figure}
\includegraphics[width=8 cm]{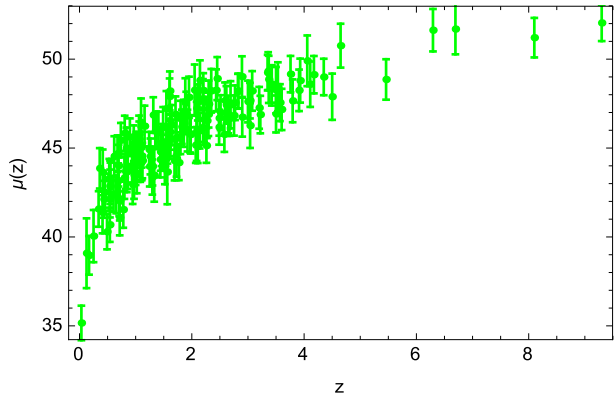}
\caption{Distance modulus $\mu(z)$ for the calibrated GRB Hubble diagram obtained by fitting
the $E_{\rm p,i}$ -- $E_{\rm iso}$ relation.}
\label{GRBHD}
\end{figure}

Gamma-ray bursts are visible up to high redshifts thanks to the
enormous energy that they release, and thus are good
candidates for our high-redshift cosmological investigation.
However, GRBs may be everything but standard candles since their
peak luminosity spans a wide range, even if there have been many
efforts to make them distance indicators using some empirical
correlations of distance-dependent quantities and rest-frame
observables \cite{Amati08}. These empirical relations allow us
to deduce the GRB rest-frame luminosity or energy from the
observer-frame measured quantity, so that the distance modulus
can be obtained with an error that depends essentially on the
intrinsic scatter of the adopted correlation. We performed our
analysis using the GRB Hubble diagram data set, built by
calibrating the $E_{\rm p,i}$ -- $E_{\rm iso}$ relation on the
Union SNIa sample \cite{MGRB1,MGRB2}.  In Table \ref{obs} we list some of the
observational data used in our analysis. \footnote{ For the full sample, please contact the authors.} After fitting the
correlation and estimating its parameters, we used it to
construct the GRB Hubble diagram up to $z\simeq 8$, which allows us to explore a very important redshift range, to determine
the expansion history of the Universe and probe properties of the dark energy.
 We recall that the luminosity
distance of a GRB with the redshift $z$ is
\begin{equation}\label{lumdist}
d_L(z) = \left( \frac{E_{\rm iso}(1 + z)}{4 \pi S_{bolo}}\right)^{1/2}.
\end{equation}
The distance modulus $\mu(z)$ is
easily obtained from the standard relation
\begin{equation}\label{modulusGBR}
\mu(z) = 5 \log_{10} { d_{L}(z, \theta_{i} )} )+\mu_0\,,
\end{equation}
where $\theta_{i}$ are model parameters and $\mu_0$ is a free parameter. The uncertainty is estimated by error
propagation.
Actually, since for GRBs the absolute calibration is not
available,  we can fit the Hubble Diagram of GRBs together with
that of SNIa and use the overlapping redshift range to
cross-calibrate the GRBs diagram, what allows to determine
$\mu_0$.
 When the correlation is fitted and its parameters are estimated, it is possible
to compute the luminosity distance of GRBs at known redshift $z$ and,
therefore, estimate the distance modulus
for each $i$\,-\,th GRB in our sample at redshift
$z_i$, and to build the Hubble diagram plotted in Fig. (\ref{GRBHD}).

\begin{table}
\centering
\resizebox{8cm}{!}{\begin{minipage}{\textwidth}
        \caption{Some data used in our analysis.}\label{obs}
 \begin{tabular}{ c|c|c|c|c||c|c}
\hline
 \multicolumn{7}{c}{Some GRBs observable quantities\footnote{It is worth noting that $E_{\rm iso}$ is not directly observable, since it depends on the cosmological model through the luminosity distance.}}    \\
\hline
redshift & $E_{\rm p,i} (keV)$ & $\sigma_{E_{\rm p,i}}$ & $S_{bolo}(10^{-6}\,erg\, cm^{-2})$ & $\sigma_{S_{bolo}}$ & $E_{\rm iso} (10^{52}\,erg )$ & $\sigma_{E_{\rm iso}}$ \\
\hline \hline\\\,
 0.03351 & 4.9 & 0.49 & 20.6219 & 2.06219 & 0.00535399 & 0.000535399 \\
 0.125 & 55. & 45. & 52.6588 & 21.0635 & 0.216774 & 0.0867097 \\
 0.1685 & 82. & 8.2 & 204.139 & 20.4139 & 1.61591 & 0.161591 \\
 0.25 & 3.37 & 1.79 & 0.127068 & 0.0317671 & 0.00244119 & 0.000610299 \\
 0.31 & 203. & 53. & 207.837 & 41.5672 & 6.55552 & 1.3111 \\
 0.3399 & 1250. & 150. & 2349.65 & 335.665 & 91.8909 & 13.1273 \\
 0.36 & 1060. & 275. & 178.062 & 48.9183 & 7.97042 & 2.18968 \\
 0.41 & 70. & 21. & 11.3826 & 2.88168 & 0.693324 & 0.175525 \\
 0.414 & 440. & 180. & 43.6518 & 8.44872 & 2.72107 & 0.526658 \\
 0.434 & 93. & 15. & 9.82091 & 1.88864 & 0.685192 & 0.131768 \\
 0.45 & 129. & 26. & 12.7073 & 1.27073 & 0.966894 & 0.0966894 \\
 0.4791 & 81.3505 & 8.13505 & 13.1412 & 1.31412 & 1.16253 & 0.116253 \\
 0.49 & 51. & 5.1 & 17.6258 & 1.8851 & 1.64622 & 0.176066 \\
 0.5295 & 61. & 15. & 1.56474 & 0.156474 & 0.176323 & 0.0176323 \\
    \hline\hline
\end{tabular}
\end{minipage}
}
\end{table}

\subsection{H(z) measurements}
The measurement of the Hubble parameter is a complementary probe
to constrain the cosmological parameters and investigate the
dark energy. The Hubble parameter can be measured using the
so-called cosmic chronometers. The most reliable cosmic
chronometers at present are old early-type galaxies that evolve
passively on a timescale much longer than their age difference,
which formed the vast majority of their stars rapidly and early
and have not experienced subsequent major episodes of star
formation or merging.  We used a list of $28$
$H(z)$ measurements obtained in this way that were compiled in \cite{farooqb}.
\subsection{Constraints from the growth rate data}
It is known \cite{Percival05} that the growth factor of density perturbations satisfies the following differential equation on subhorizon scales ($k^2\gg
a^2H^2$), where primes denote differentiation with respect to the scale factor $a$
\begin{equation}
\delta''(a)+\left(\frac{3}{a}+\frac{H'(a)}{H(a)}\right)\delta'(a)
-\frac{3}{2}\frac{\omms }{a^5 H(a)^2/H_0^2}\,\delta(a)=0\,,
\label{ode}
\end{equation}
where $\delta(a)=\frac{\delta\rho_m}{\rho_m}$ denotes the cosmological  matter overdensity \footnote{ It is
worth noting that this equation  is not  valid for a scalar tensor theory,  since in this theory the dark energy is partially gravitationally clustered even at small scales. However, the correction is small in the GPS model \cite{BEFPS00}.}. This differential equation
has in general two solutions that correspond to two modes, a growing and a decaying one, that in a matter dominated
universe scale as $\delta=a$ and as $\delta=a^{-3/2}$ respectively. In order to numerically integrate the Eq.
(\ref{ode}), we set the usual initial conditions: $\delta(a_{in})\simeq a_{in}$ and $\delta'(a_{in})\simeq1$. The
growth rate is defined as $f(a)=\displaystyle{\frac{d  \delta(a)}{d \log a}}$. Most of the growth rate data are provided
by peculiar velocity measurements in galaxy surveys and are obtained in terms of galaxy density, which is related to
the matter perturbation by the relation $\delta_g=b \delta_m$, where $b$ is the so called (unknown) bias parameter.
Therefore  measurements of $f$ depend on the value of the bias parameter. A more reliable function is the
product $f(z)\sigma_8(z)=f\sigma_8(z)$ where $\sigma_{8}(z)$ is the amplitude of the power spectrum of density perturbations on the scale
$8h^{-1}$Mpc, as it is independent on the bias and can be measured also from
weak lensing surveys. Here we use  the Gold-2017 compilation of 18 $f\sigma_8(z)$ measurements, presented in \cite{Nesseris2017}.
It is worth noting that to use the Gold-2017 growth data we follow the same procedure as explained in \cite{Nesseris2017} to
correct the data for the Alcock-Paczynski effect mentioned in  that
 paper. Actually the  growth rate data depend on the fiducial model used  to convert redshifts to distances. Following
 \cite{Nesseris2017}  we rescaled the measurements by the ratios of $H(z)d_A(z) $ of our model to that of the fiducial one.

\section{Statistical analysis}
\label{stat}
\begin{figure}
\includegraphics[width=8 cm]{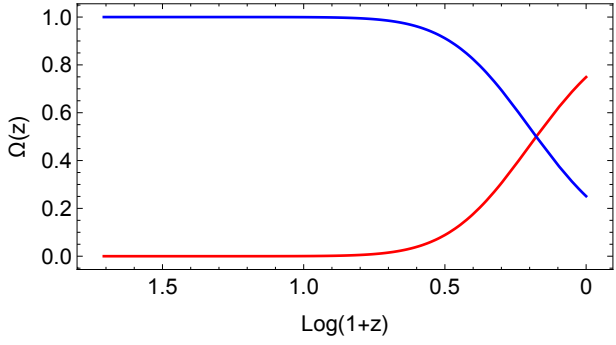}
\caption{Redshift evolution of the $\Omega$s parameters for the GPS model, corresponding to the  best fit values of model parameters. The blue line represents $\Omega_m(z)$, and the red line  $\Omega_{GPS}(z)$}
\label{omegas}
\end{figure}

\begin{figure}
\includegraphics[width=8 cm]{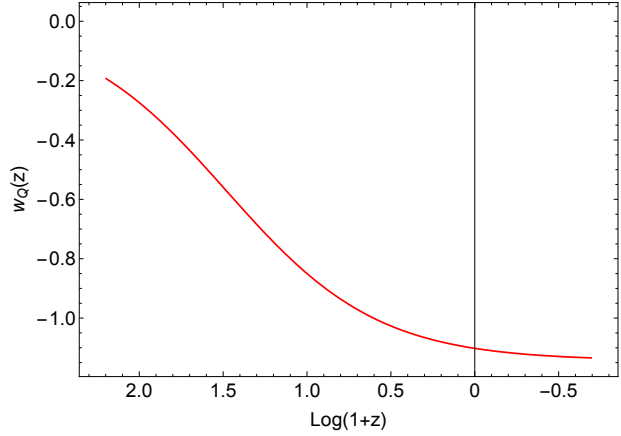}
\caption{Redshift evolution of the  equation of state for the $GPS$ model, corresponding to the best fit values
of  model parameters, when the full dataset is used. } \label{wcompare}
\end{figure}

To constrain the parameters of the GPS dark energy model we performed a preliminary and standard fitting procedure
to maximize the likelihood function ${\cal{L}}({\bf p})$. This requires the knowledge of the precision matrix,
that is, the inverse of the covariance matrix of the measurements,

\begin{eqnarray}
\footnotesize
{\cal{L}}({\bf p}) & \propto & \frac {\exp{(-\chi^2_{SNIa/GRB}/2)}}{(2 \pi)^{\frac{{\cal{N}}_{SNIa/GRB}}{2}} |{\bf C}_{SNIa/GRB}|^{1/2}} \frac{\exp{(-\chi^2_{H}/2})}{(2 \pi)^{{\cal{N}}_{H}/2} |{\bf C}_{H}|^{1/2}}\nonumber\\&& \frac {\exp{(-\chi^2_{f\sigma_8}/2)}}{(2 \pi)^{\frac{{\cal{N}}_{f\sigma_8}}{2}} |{\bf C}_{f\sigma_8}|^{1/2}}  \nonumber \,,
\label{defchiall}
\end{eqnarray}

where

\begin{equation}
\chi^2(\bf p) = \sum_{i,j=1}^{N} \left( \data_i - x^{th}_i(p)\right)C^{-1}_{ij} \left( \data_j - x^{th}_j(p)\right) \,.
\label{eq:chisq}
\end{equation}
Here $\bf p$ is the set of parameters, $N$ is the number of data points, $\bf \data_i$ is the $i-th$ measurement; $\bf
x^{th}_i(p)$ indicates the theoretical prediction for this measurement, it depends on the parameters $\bf p$;, $\bf
C_{ij}$ is the covariance matrix (specifically,\\ $ {\bf C}_{SNIa/GRB/H/f\sigma_8}$ indicates the SNIa/GRBs/H/$f\sigma_8$
covariance matrix).   Moreover we use a gaussian prior term $\frac{1}{\sqrt{2 \pi \sigma^2}} \exp{\left[ - \frac{1}{2} \left ( \frac{h - h_{peak}}{\sigma} \right )^2\right]} $, where \cite{shoes}
\begin{eqnarray}
&& h_{peak} =h_{shoes}\,, \\
&& \sigma = 5\sigma_{h_{shoes}}\,.\nonumber
\end{eqnarray}
 It is worth to stress that in  Tables (\ref{tab1b}) and (\ref{tab2b}) the inferred value of $h$ is strongly influenced by this prior, because neither SNIa, nor GRB, nor $f\sigma_8(z)$ data sets by itself do not allow us to determine $h$. To sample the ${\cal{N}} $ dimensional space of parameters, we use the MCMC
method and ran three parallel chains and use the Gelman - Rubin diagnostic approach to test the convergence. As a test
probe, it uses the reduction factor $R$, which is the square root of the ratio of the variance between-chains and the
variance within-chains. A large $R$ indicates that the variance between-chains is substantially greater than the
variance within-chain, so that a longer simulation is needed. We require that $R$ converges to 1 for each parameter. We
set $R - 1$ to be of order $0.1$. We discarded the first $30\%$ of the point iterations at the beginning of any MCMC run, and
thinned the chains that were run many times. We finally extracted the constrains on the parameters by coadding the
thinned chains. The histograms of the parameters from the merged chains were then used to infer median values and
confidence ranges: the $15.87$th and $84.13$th quantiles define the $68 \%$ confidence interval; the $2.28$th and
$97.72$th quantiles define the $95\%$ confidence interval; and the $0.13$th and $99.87$th quantiles define the $99\%$
confidence interval. In Tables (\ref{tab1}), (\ref{tab1b}) we present results of our analysis. In Figs. (\ref{wcompare}) and (\ref{omegas}) we plot respectively the behaviour
of the GPS equation of state and the $\Omega$s parameters corresponding to the best fit values of the parameters.

\begin{table*}
\begin{center}
\resizebox{11cm}{!}{
\begin{tabular}{cccccccccc}
\hline
\, & \multicolumn{7}{c}{\bf GPS Dark energy}   \\
\, & \, & \, & \, & \, & \, & \, & \, & \,   \\
\hline
\, & \, & \, & \, & \, & \, & \, & \, & \,   \\
$Id$ & $\langle x \rangle$ & $\tilde{x}$ & $68\% \ {\rm CL}$  & $95\% \ {\rm CL}$ &  $\langle x \rangle$ & $\tilde{x}$ & $68\% \ {\rm CL}$  & $95\% \ {\rm CL}$ \\
\hline \hline
\, & \, & \, & \, & \, & \, & \, & \, & \,   \\
\hline \, & \multicolumn{4}{c}{SNeIa/GRBs/H(z)/ $f\sigma_8(z)$}  \, & \multicolumn{4}{c}{SNeIa/H(z)/$f\sigma_8(z)$}
 \\
\hline
\, & \, & \, & \, & \, & \, & \, & \, & \,   \\
$\Omega_m$ &0.27 &0.27& (0.25, 0.31) & (0.22, 0.32) &0. 295&0.3& (0.28, 0.32) & (0.25, 0.34)  \\
\, & \, & \, & \, & \, & \, & \, & \, & \,  \\
$w_0$ &-1.13& -1.14& (-1.3, -0.96) & (-1.4,  -0.78)  &-0.99& -0.98& (-1.12, -0.84) & (-1.27, -0..73) \\
\, & \, & \, & \, & \, & \, & \, & \, & \, &  \\
$\alpha$ &3.0&2.8 & (2.1, 4.2) & (2.05, 4.8) &2.9& 2.8& (-2.17, 3.5) & (2.01, 4.4)  \\
\, & \, & \, & \, & \, & \, & \, & \, & \, &  \\
$\beta$ &0.07&0.08& (0.03, 0.11) & (0.02, 0.16) &0.08& 0.08& (0.05,0.11) & (0.02, 0.16)  \\
\, & \, & \, & \, & \, & \, & \, & \, & \, &  \\
$h$ &0.70& 0.7 & (0.69, 0.71) & (0.67, 0.72)  &0.71& 0.71 & (0.69, 0.71) & (0.68, 0.73)  \\
\, & \, & \, & \, & \, & \, & \, & \, & \, &  \\
\hline
\end{tabular}}
\end{center}
\caption{Constraints on the GPS parameters from different data: combined SNIa and  GRBs Hubble diagrams, $f\sigma_8(z)$ data sets and
$H(z)$ data sets (Left Panel); and SNIa  Hubble diagram, $f\sigma_8(z)$ data sets and
$H(z)$ data sets (Right Panel ). Columns report the mean $\langle x \rangle$ and median $\tilde{x}$ values  and the $68\%$ and $95\%$
confidence limits. } \label{tab1}
\end{table*}

\section{Confrontation of the GPS model with the cosmological constant model of dark energy}
\begin{table*}
\begin{center}
\resizebox{9 cm}{!}{
\begin{tabular}{cccccccccc}
\hline
\, & \multicolumn{4}{c}{\bf GPS Dark energy}   \\
\, & \, & \, & \, & \, & \, & \, & \, & \,   \\
\hline
\, & \, & \, & \, & \, & \, & \, & \, & \,   \\
$Id$ & $\langle x \rangle$ & $\tilde{x}$ & $68\% \ {\rm CL}$  & $95\% \ {\rm CL}$ \\
\hline \hline
\, & \, & \, & \, & \, & \, & \, & \, & \,   \\
\hline \, & \multicolumn{4}{c}{SNeIa/GRBs/ $f\sigma_8(z)$}
 \\
\hline
\, & \, & \, & \, & \, & \, & \, & \, & \,   \\
$\Omega_m$ &0.20 &0.21& (0.17, 0.23) &(0.15, 0.31) \\
\, & \, & \, & \, & \, & \, & \, & \, & \,  \\
$w_0$ &-1.1& -1.14& (-1.2, -0.97) & (-1.4,  -0.88)  \\
\, & \, & \, & \, & \, & \, & \, & \, & \, &  \\
$\alpha$ &2.8&2.7 & (2.1, 3.6) & (2.02, 4.2) \\
\, & \, & \, & \, & \, & \, & \, & \, & \, &  \\
$\beta$ &0.04&0.03& (0.02, 0.06) & (0.02, 0.1)  \\
\, & \, & \, & \, & \, & \, & \, & \, & \, &  \\
$h$ &0.69& 0.69 & (0.69, 0.7) & (0.68, 0.72)  \\
\, & \, & \, & \, & \, & \, & \, & \, & \, &  \\
\hline
\end{tabular}}
\end{center}
\caption{Constraints on the GPS parameters from SNIa and  GRBs Hubble diagrams, and $f\sigma_8(z)$ data sets.  Columns report the mean $\langle x \rangle$ and median $\tilde{x}$ values  and the $68\%$ and $95\%$
confidence limits. } \label{tab1b}
\end{table*}

It is interesting to compare predictions of the GPS model of
dark energy with predictions of the Standard $\Lambda$CDM  model that relates the observed
accelerated expansion of the Universe to the non-zero value of the cosmological
constant.

When the accelerated expansion of the Universe is due to the
cosmological constant the dark energy equation of state is
$p_{DE} = -\varrho_{DE}$, so $w = -1$. From the continuity
equation (4) it follows that $\varrho_{DE} = {\rm const}$. In
this case at the post-recombination epoch assuming spatially
flat $\Lambda$CDM model we have
\begin{equation}
H^{2}(z) = H^{2}_{0}\big( \Omega_{m}(1 + z)^3 + 1 -
\Omega_{m}\big)\,. \label{hubblambda}
\end{equation}
Using this Hubble expansion rate the luminosity distance
$d_{L}$ and the angular diameter distance $d_{A}(z)$ are  defined as
\begin{eqnarray}
d_L(z) &=&\frac{c}{H_0} (1+z)\int_0^z \frac{dy}{ \sqrt{1-\Omega_m +\Omega_m (y+1)^3}},\label{lumd} \\
d_A(z) &=&\frac{c}{H_0}\frac{1}{1+z} \int_0^z \frac{dy}{ \sqrt{1-\Omega_m +\Omega_m (y+1)^3}},\label{angd}
\end{eqnarray}
Using this luminosity distance, we can evaluate the distance
modulus from its standard definition\,
\begin{equation}
\mu(z) = 25 + 5 \log{d_L(z)}\,. \label{eq:defmu}
\end{equation}

 In Table  (\ref{tab2}) and (\ref{tab2b}) we present results of the same
statistical analysis as performed for the GPS model for the $\Lambda$CDM model, using the same data sets. It turns out  that $\mu^{GPS}_0\simeq 0.68$, and  $\mu^{\Lambda CDM}_0\simeq0.7$.
\begin{table*}
\begin{center}
\resizebox{11cm}{!}{
\begin{tabular}{cccccccccc}
\hline
\, & \multicolumn{7}{c}{ \bf $\Lambda$CDM}   \\
\, & \, & \, & \, & \, & \, & \, & \, & \,   \\
\hline
\, & \, & \, & \, & \, & \, & \, & \, & \,   \\
$Id$ & $\langle x \rangle$ & $\tilde{x}$ & $68\% \ {\rm CL}$  & $95\% \ {\rm CL}$ &  $\langle x \rangle$ & $\tilde{x}$ & $68\% \ {\rm CL}$  & $95\% \ {\rm CL}$ \\
\hline \hline
\, & \, & \, & \, & \, & \, & \, & \, & \,   \\
\hline \, & \multicolumn{4}{c}{SNeIa/GRBs/H(z)/$f\sigma_8(z)$ }  \, & \multicolumn{4}{c}{SNeIa/H(z)/$f\sigma_8(z)$}
 \\
\hline
\, & \, & \, & \, & \, & \, & \, & \, & \,   \\
$\Omega_m$ &0.26 &0.26& (0.24, 0.28) & (0.22, 0.3) &0. 25&0.25& (0.25, 0.27) & (0.23, 0.32)  \\
\, & \, & \, & \, & \, & \, & \, & \, & \,  \\
$h$ &0.70& 0.70 & (0.69, 0.71) & (0.68, 0.72)  &0.72& 0.72 & (0.69, 0.73) & (0.68, 0.74)  \\
\, & \, & \, & \, & \, & \, & \, & \, & \, &  \\
\hline
\end{tabular}}
\end{center}
\caption{Constraints on the $\Lambda$CDM parameters from different data: combined SNIa and  GRBs Hubble diagrams, $f\sigma_8(z)$ data sets and
$H(z)$ data sets (Left Panel); and SNIa  Hubble diagram, $f\sigma_8(z)$ data sets and
$H(z)$ data sets (Right Panel). Columns report the mean $\langle x \rangle$ and median $\tilde{x}$ values  and the $68\%$ and $95\%$
confidence limits. } \label{tab2}
\end{table*}

\begin{table*}
\begin{center}
\resizebox{8 cm}{!}{
\begin{tabular}{cccccccccc}
\hline
\, & \multicolumn{4}{c}{\bf $\Lambda$CDM}   \\
\, & \, & \, & \, & \, & \, & \, & \, & \,   \\
\hline
\, & \, & \, & \, & \, & \, & \, & \, & \,   \\
$Id$ & $\langle x \rangle$ & $\tilde{x}$ & $68\% \ {\rm CL}$  & $95\% \ {\rm CL}$ \\
\hline \hline
\, & \, & \, & \, & \, & \, & \, & \, & \,   \\
\hline \, & \multicolumn{4}{c}{SNeIa/GRBs/ $f\sigma_8(z)$}
 \\
\hline
\, & \, & \, & \, & \, & \, & \, & \, & \,   \\
$\Omega_m$ &0.23 &0.23& (0.19, 0.26) &(0.17, 0.29) \\
\, & \, & \, & \, & \, & \, & \, & \, & \,  \\
$h$ &0.69& 0.7& (0.69, 0.7) & (0.67, 0.73)  \\
\, & \, & \, & \, & \, & \, & \, & \, & \, &  \\
\hline
\end{tabular}}
\end{center}
\caption{Constraints on the $\Lambda$CDM parameters from SNIa and  GRBs Hubble diagrams, and $f\sigma_8(z)$ data sets.  Columns report the mean $\langle x \rangle$ and median $\tilde{x}$ values  and the $68\%$ and $95\%$
confidence limits. } \label{tab2b}
\end{table*}

\section{Discussion of our calibration procedure of the $E_{\rm p,i}$ -- $E_{\rm iso}$ relation }
In this section we discuss the reliability, for cosmological applications, of our calibration technique
of the $E_{\rm p,i}$ -- $E_{\rm iso}$ relation, based on Type Ia supernovae Hubble diagram. We are specially interested in understanding
how much this calibration procedure affects the independence of the (SNIa and GRBs) datasets. We  already discussed this topic in some
previous papers (see \cite{MGRB1} and references therein).  However, in order to further investigate this question we performed an independent
calibration,  based on an approximate formula for the luminosity distance which holds in any cosmological model, and not on a power
series expansion in the redshift parameter z, as in the cosmographic approach. Our starting point is the well known relation between
the angular diameter distance $d_A$ and the luminosity distance $d_L$
\begin{equation}\label{lum2}
d_{L}=\left(1+z\right)^2d_A,
\end{equation}
where the angular diameter distance $d_A$ is a solution of the equation
\begin{equation}   \left ( \frac{dz}{dv}
\right )^2 \frac{d^2 d_A}{dz^2} + \left (   \frac{d^2z}{dv^2} \right )
\frac{dd_A}{dz} +\frac{4\pi G}{c^{4}} T_{\alpha   \beta} k^{\alpha} k^{\beta} d_A = 0.
\label{eq:angdiam2}
\end{equation}
with the following initial conditions:
\begin{eqnarray}\label{eq:cauchy}   &&d_A(z)\arrowvert_{z = 0} = 0,
\nonumber\\ && \\ &&\frac{dd_A(z)}{dz}   \arrowvert_{z = 0} =
\frac{c}{H_0}.\nonumber
\end{eqnarray}
In Eq. (\ref{eq:angdiam2}) $\it{v}$ is the affine parameter, $T_{\alpha   \beta}$ is the matter density tensor,
$k^{\alpha}=\displaystyle{{dx^{\alpha}\over dv}}=-g^{\alpha \beta}\Sigma,_{\beta}$ is the vector field tangent to the congruence of light rays,
and $\Sigma$ is the null surface along which the light rays propagate from the source.
In the general form this equation is very
complicated (\cite{Kant98,Kant20,Kant01,approx}), and in most cases it does not admit analytical solution. From
the mathematical point of view it turns out that it is of Fuchsian type with several regular singular points and
a regular singular point at infinity.  When we introduced the dimensionless angular diameter distance
 $r=d_A H_{0}/ c$ we discovered  \cite{approx} that there is a simple function, which quite accurately
reproduces the exact numerical solutions of the equation (\ref{eq:angdiam2})  for $z$ up to  high values.
Here we generalize this function, to extend the accuracy of this approximation up to $z\simeq 10$. This function $r(z)$ has the form
\begin{equation}
r(z)=\frac{z (z+1)^2}{\sqrt{d_3 z^3+\left(d_2 z^2+d_1 z+1\right){}^2}}\,.,
\label{eq:approxnew}
\end{equation}
where $d_1$, $d_2$ and $d_3$ are constants that depend on parameters of the considered cosmological model.
Moreover the function (\ref{eq:approxnew}) automatically
satisfies the imposed initial conditions, so $r(0)=0$ and
$\displaystyle{dr \over dz}(0)=1$.

This approximate expression immediately provides an empirical formula for the approximate luminosity distance.
The GRBs and the SNIa samples have been fitted simultaneously with this approximated luminosity function.
For what it concerns the GRBs sample, our task is to determine the parameters $\{ a, b, d_1, d_2,d_3\} $.
Actually the $E_{\rm p,i}$ -- $E_{\rm iso}$ relation can be written in the form
\begin{equation}
\log_{10} S_{\rm bol} = a + b \log_{10} E_{\rm p,i} - \log_{10}[4\pi {d^2_{L\,approxi}}(z, d_1, d_2, d_3)].
\label{eqfull}
\end{equation}
To efficiently sample the $5$-dimensional parameter space, we used the MCMC method and ran three
parallel chains and used the Gelman-Rubin convergence test, as described in the previous section. It turns out that
the calibration parameters $a$, $b$, and $\sigma$ are fully consistent with the results obtained from the  SNIa
sample based calibration (see \cite{MGRB1})\,, confirming the reliability of our calibration technique\,: we actually
find that\footnote{$\sigma_{\rm int}$ is the intrinsic dispersion, characterizing the $E_{\rm p,i}$ -- $E_{\rm iso}$ relation\,\cite{MGRB2}.}$a= 1.87\pm 0.09$\, $\sigma_{\rm int}=0.36_{-0.02}^{+0.03}$\,  $b=52.5_{-0.1}^{+0.13}$,  $d_1=1.17^{+0.11}_{0.1}$, $d_2=0.33^{+0.04}_{-0.05}$ and $d_3=0.2^{+0.2}_{-0.1}$.
It is worth noticing that the Eq. (\ref{eqfull}) can be used also in a full Bayesian procedure to estimate the cosmological parameters
and the additional parameters of the $E_{\rm p,i}$ -- $E_{\rm iso}$ correlation \cite{MGRB1, MEC11}. However it is clear that in this case the values
of the correlation parameters, which are important for cosmological applications, unfortunately depend on the assumed background cosmological model, so they do not provide an independent calibration.

\section{Comparison of the GPS model with the $\Lambda$CDM model }
To compare the different models presented in the previous sections with the data and to check if we can discriminate
them, we use the Akaike Information Criterion (AIC)  \cite{aic} \cite{aic2}, and its indicator
\begin{equation} \label{aic}
AIC=-2 \ln{\cal{L}_{\bf max} }+  2 k_p +\frac{2 k_p (k_p+1)}{N_{tot}-k_p-1}\,,
\end{equation}
where $N_{tot} $ is the  total number of data and $k_p$ the number of free parameters (of the cosmological model).
In our case we have $N_{tot}=815$, and $k_p=6$. It turns out that the smaller is the value of AIC the better is  the fit to  the data. To compare different cosmological models  we introduce the model  difference
$ \Delta_{AIC} = AIC_{model} -AIC_{min}$. The relative difference corresponds to different cases:
$4 < \Delta_{AIC} < 7$ indicate a positive evidence against the model with higher value of $AIC_{model}$,
while $\Delta_{AIC} \geq 10 $ indicate a strong evidence. If, $ \Delta_{AIC} \leq 2$ is an indication that
the two  models are consistent. In our case we have found that  the model with the lower  AIC is the GPS model
and  $ \Delta_{AIC} = 0.7 $ if we consider GRBs, and  $ \Delta_{AIC} = 5.7 $ without GRBS. This result
indicates that the two models are statistically consistent, if we consider GRBs data, and that the $\Lambda$CDM
model would be slightly favoured  without GRBs.

\section{Conclusions}
We have compared two different models of dark energy with presently available observational data. We show that with
appropriate choice of the parameters of these models they are compatible with observations.  Our statistical analysis
indicates that the GPS model seems to be slightly more favoured than the $\Lambda$CDM model, if we consider the GRBs Hubble diagram.
It means that to further restrict
different models of dark energy it will be necessary to increase the precision of the  Hubble diagram at high redshifts,
and  to perform more detailed analysis of the influence of dark energy on the process of formation of large scale
structure and in particular on its late evolution at $z < 2$. Of course, more and more precise observational data will
reduce statistical errors and could lead to further restrictions on parameters describing properties of dark energy and
better differentiate different models.

\section{Acknowledgments}
MD is grateful to the INFN for financial support through the Fondi FAI GrIV. EP acknowledges the support of INFN Sez.
di Napoli  (Iniziativa Specifica QGSKY).

\end{document}